\documentclass[reprint,aip,nofootinbib,superscriptaddress]{revtex4-1}
\usepackage{amsmath}
\usepackage{amssymb}
\usepackage{cancel}
\usepackage{amsfonts}
\usepackage[T1]{fontenc}
\usepackage{bm}
\usepackage{color}
\usepackage{graphicx}
\DeclareSymbolFont{operators}{OT1}{cmr}{m}{n}
\DeclareSymbolFont{letters}{OML}{cmm}{m}{it}
\DeclareSymbolFont{symbols}{OMS}{cmsy}{m}{n}
\DeclareSymbolFont{largesymbols}{OMX}{cmex}{m}{n}

\usepackage{hyperref}
\hypersetup{
    bookmarks=true,         
    unicode=false,          
    pdftoolbar=true,        
    pdfmenubar=true,        
    pdffitwindow=false,     
    pdfstartview={FitH},    
    pdfsubject={Plasma fluid theory},   
    pdfproducer={GNU Make}, 
    pdfkeywords={keyword1} {key2} {key3}, 
    pdfnewwindow=true,      
    colorlinks=true,        
    linkcolor=blue,      
    citecolor=blue,         
    filecolor=blue,      
    urlcolor=blue           
}
\usepackage{array}

\newcommand{\1}{\hat{\bm{1}}}

\makeatother

\begin{document}

\title{Relativistic-Amplitude electromagnetic waves -- Beating the ``magnetic'' barrier}

\author{Swadesh Mahajan}
\email{mahajan@mail.utexas.edu}
\affiliation{Institute for Fusion Studies, University of Texas at Austin, Austin, TX 78712, USA}
\affiliation{Department of Physics, School of Natural Sciences, Shiv Nadar University, Uttar Pradesh 201314, India}
\author{Manasvi Lingam}
\email{manasvi.lingam@cfa.harvard.edu}
\affiliation{Institute for Theory and Computation, Harvard-Smithsonian Center for Astrophysics, Cambridge, MA 02138, USA}
\affiliation{John A. Paulson School of Engineering and Applied Sciences, Harvard University, Cambridge, MA 02138, USA}

\begin{abstract}
The dispersion characteristics of an circularly polarized electromagnetic wave of arbitrary amplitude, propagating in a highly (thermally and kinematically) relativistic plasma, are shown to approach those of a linear wave in an unmagnetized, non-relativistic plasma. Further aided by high relativistic temperatures, the cut-off frequency tends to become negligibly small; as a result, waves with frequencies well below the nominal plasma and the cyclotron frequencies find the plasma to be essentially transparent. This relativistic phenomenon may greatly advance our ability to understand and model the dynamics of a large class of astrophysical and laser-produced high energy density systems.

\end{abstract}

 
\maketitle

\section{Introduction}
Uncovering the dynamics of relativistic plasmas \citep{Lich67,Anil05,RZ13} plays a key role in defining and determining the characteristics of a variety of astrophysical systems and phenomena. Examples of relativistic plasmas include the early Universe \citep{BM94,Sub16}, relativistic jets \citep{MR99,KZ15}, accretion discs and flows around black holes \citep{TNM11,YN14}, gamma ray bursts \citep{Pir04,KZ15}, pulsar magnetospheres \citep{KT73,Mich82,Spit06}, Active Galactic Nuclei \citep{Anto93}, supernovae and collapsars \citep{WB06} whereas turbulence and magnetic reconnection \citep{BH98,Bis03,ZY09,NR11,CLHB} constitute some of the astrophysical processes of interest. Relativistic plasma dynamics also plays a key role in modeling high energy density laboratory systems typically created by lasers such as inertial containment fusion \citep{Rot01,Lind04,Glen10}, general laser-plasma interactions \citep{MTB06,MBP13,RD16}, and laboratory astrophysics \citep{RDR06,Bul15}. 

Since  waves (collective oscillations) are, arguably, the most evident expression of plasma dynamics, studying their behavior in a relativistic plasma is undoubtedly a prime theoretical necessity. In addition, waves also have a wide variety of applications in astrophysical, space and fusion plasma environments \citep{Stix62,Whit,Cran01}.

Because of the nonlinear relationship between the momentum and kinematic velocity,
\begin{equation} \label{RelGamma}
\boldsymbol P= m \gamma \boldsymbol V = \frac {m \boldsymbol V}{\sqrt{1-V^2}},
\end{equation}
via the kinematic $\gamma=(\sqrt{1-V^2})^{-1}$, the relativistic dynamics tends to be nonlinear even when the equivalent non-relativistic (N.R) dynamics, obeying the simpler relation $\boldsymbol P_{N.R} = m \boldsymbol V$, may be linear. Consequently, an investigation of relativistic phenomena - for instance, the relativistic equivalents of the familiar linear plasma waves - can pose new challenges.

Despite these intrinsic difficulties, we demonstrate, in this paper, that a fully relativistic two-fluid system can support arbitrary amplitude waves whose ``dispersion'' displays strikingly different properties as compared to their well-known N.R limits. More importantly, these new features may help us understand and better interpret the nature of waves and turbulence in highly relativistic astrophysical and laboratory systems like the ones listed above. 

\section{Mathematical preliminaries}
We will analyze the simplest non-trivial two-fluid plasma consisting of oppositely charged particle species with rest frame masses $m_\pm$, and charges $q_\pm$. The plasma  is taken to be quasi-neutral (in the rest frame, $n_{R+} = n_0 = n_{R-}$), and is allowed to be relativistic, kinematically as well as thermally. Here, we wish to emphasize that the model studied herein contains relativistic MHD \citep{MM03,Anil05,RZ13} as a limiting case \citep{Mah03,ZHK09,AC15}. Our results encompass, to varying degrees, some of the previous analyses of linear and non-linear relativistic magnetofluid waves \citep{KePe76,SaKa80,SK80,GLST85,SRY,AMV09,SMH10,LAM12,Kaw17}. 

For simplicity, constant but arbitrary temperatures for the two species are assumed. Notice that assigning a temperature is essential to reflect the spread in the particle energy spectrum discussed in the standard model for the pulsar magnetosphere \citep{GJ69}. For such a system, the dynamical equations may be written in the compact form
\begin{equation} \label{GVEq}
\frac{\partial \boldsymbol{\Omega}_\pm}{\partial t} =  \nabla \times \left({\bf V}_\pm \times \boldsymbol{\Omega}_\pm\right),
\end{equation}
in terms of the generalized vorticity,
\begin{equation} \label{GVDef}
\boldsymbol{\Omega}_\pm = {\bf B} \pm \mu_\pm \nabla \times \left(\gamma_\pm {\bf V}_\pm\right)={\bf B} \pm \mu_\pm \nabla\times {\bf U}_\pm,
\end{equation}
where ${\bf U}_\pm= \gamma_\pm {\bf V}_\pm$ is the spatial four-velocity, and $\mu_\pm=m_\pm f_\pm/ (m_{+} f_{+} + m_{-}f_{-})$ are the effective mass factors, i.e, the masses enhanced by the thermal factor $f >1$ \citep{Bek87,Mah03}. The factor $f$ acts like a ``thermal $\Gamma$'' factor and becomes unity for N.R temperatures; by definition $\mu_+ + \mu_- = 1$. 

An important caveat worth mentioning here is that our model does not include kinetic effects (e.g. FLR contributions) and dissipative terms such as the plasma resistivity \citep{HM02,THM08,LHP17}; we plan to incorporate these terms and study their effects on relativistic plasma waves in a subsequent publication. For example, in the non-relativistic limit, the inclusion of FLR terms could lead to the standard Alfv\'en wave dispersion relation being replaced by the kinetic Alfv\'en wave. In astrophysical environments with high magnetic fields, such as pulsar magnetospheres, we might expect FLR effects to be relatively unimportant provided that the condition $k \rho_i \ll 1$ is valid, where $\rho_i$ is the ion gyroradius.  

Equations (\ref{GVEq}) and (\ref{GVDef}) are \emph{dimensionless}; the respective normalizations are: the magnetic field by a fiducial value $B_0$ (as measured in the rest frame), velocities to $c$, and time (length) to $\Omega^{-1}_0$ $(c/\Omega_{0})$, where $\Omega_0 = |q| B_0/(m_{+} f_{+} + m_{-}f_{-}) c$ is some representative measure of the cyclotron frequency in the ambient fiducial field. This set of normalizing factors converts Maxwell's equation to the dimensionless form 
\begin{equation} \label{MaxEqn}
\nabla \times \left(\nabla \times {\bf B}\right) + \frac{\partial^2 {\bf B}}{\partial t^2} = \left(\frac{1}{{U}_{A0}}\right)^2 \nabla \times \left( {\bf U}_+ - {\bf U}_-\right) 
\end{equation}
where $U_{A0}= B_0/c\sqrt{4\pi n_{0}(m_{+} f_{+} + m_{-}f_{-})}$ can be viewed as a corresponding normalized Alfv\'en speed.

Following the standard procedure for plasma waves, we assume that the plasma is embedded in a constant magnetic field of strength $B_0$. By orienting the $z$-axis along this ambient field, the normalized magnetic field in the preceding equations is expressible as
\begin{equation} \label{AmbB}
{\bf B}= \hat{e}_z + {\bf b}
\end{equation}
where ${\bf b}$ is the space-time dependent dynamic (wave) field. The fluid velocity ${\bf V}$ can also be decomposed into an ambient and a dynamic part, 
 \begin{equation} \label{AmbV}
{\bf V}=  V_0(z) \hat{e}_z + {\bf v}
\end{equation}
where ${\bf v}$ is the velocity associated with the wave component. Unlike linear waves, we shall allow $|{\bf b}|$ to become arbitrary large, and  $|\bf v|$ to approach unity; hence, the relativistic wave amplitudes can be as large as necessary.

If $V_0$ were constant, then its effects are somewhat trivial (one could just work in a frame where $V_0=0$). However, a spatially varying $V_0$ (a sheared flow) is rather interesting; the free energy available in the sheared flow could amplify the waves we are about to investigate \citep{MMR97,PRM98}. We shall first begin by deriving the relativistic nonlinear waves for the case $V_0=0$. 

\section{Zero background flow}
Splitting $\bf B$ as in (\ref{AmbB}), and ${\bf V}_\pm = {\bf v}_\pm$, the equations of motion (\ref{GVEq}) and (\ref{GVDef}), and the Maxwell equation (\ref{MaxEqn}) transform into 
\begin{eqnarray} \label{GVEqDec}
\frac{\partial \left[{\bf b} \pm \mu_\pm \nabla \times \left(\gamma_\pm {\bf v}_\pm\right) \right]}{\partial t} =  \nabla \times \left({\bf v}_\pm \times \hat{e}_z\right) && \\
 + \nabla \times \left({\bf v}_\pm \times \left[{\bf b} \pm \mu_\pm \nabla \times \left(\gamma_\pm {\bf v}_\pm\right) \right]\right) && \nonumber,
\end{eqnarray}
\begin{equation} \label{MaxEqn2}
\nabla \times \left(\nabla \times {\bf b}\right) + \frac{\partial^2 {\bf b}}{\partial t^2} = \left(\frac{1}{U_{A0}}\right)^2 \nabla \times \left( \gamma_{+}{\bf v}_+ - \gamma_{-}{\bf v}_-\right).
\end{equation}
These equations are intrinsically nonlinear, not only through the terms that involve ``products'' of the wave fields, but also through $\gamma_{\pm}= (1/\sqrt {1-\bf v_{\pm}\cdot \bf v_{\pm}})$. 

It is easy to verify that the $\gamma$ factor becomes independent of space-time if the velocity field is circularly polarized (CP). We thus seek circularly polarized, plane wave solutions  
\begin{equation}\label{Circv}
{\bf v}_\pm = \frac{v_\pm}{2} \left[v_\pm \left(\hat{e}_x + i \hat{e}_y\right) e^{i(kz-\omega t)} + \mathrm{c.c}\right],
\end{equation}
\begin{equation}\label{Circb}
{\bf b} = \frac{b}{2} \left[\left(\hat{e}_x + i \hat{e}_y\right) e^{i(kz-\omega t)} + \mathrm{c.c}\right],
\end{equation}
where $b$ and $v_\pm$ are constant (albeit arbitrary magnitude) amplitudes of the wave of frequency $\omega$ propagating (with a wave number $k$) along the ambient magnetic field. The relativistic factors
\begin{equation} \label{gamma}
\gamma_{\pm}= \frac{1}{\sqrt {1-\bf v_{\pm}\cdot \bf v_{\pm}}}= \frac{1}{\sqrt {1-v^2_{\pm}}},
\end{equation}
now depend only on the constant wave amplitudes, and can thus be treated as numbers henceforth. The circularly polarized ansatz facilitates another simplification simultaneously; it fully eliminates the most complicated nonlinear term, namely, the second term on the right hand side of (\ref{GVEqDec}). This happens because, for the CP fields, 
\begin{eqnarray}
&& \hat{e}_z \times {\bf v}_\pm = -i {\bf v}_\pm, \quad \hat{e}_z \times {\bf b} = -i {\bf b}, \nonumber \\
&& \nabla \times {\bf v}_\pm = k {\bf v}_\pm, \quad \nabla \times {\bf b} = k {\bf b}, \nonumber \\
&& {\bf b} \times {\bf v}_\pm = 0, \quad {\bf b} \times \left(\nabla \times {\bf v}_\pm\right) = 0, \nonumber
\end{eqnarray}
making the nonlinear term identically zero. Following this drastic simplification, and using Eqs. (\ref{Circv})
and (\ref{Circb}), the system (\ref{GVEqDec}) and (\ref{MaxEqn2}) reduces to three-coupled  algebraic (nonlinear) relations between the amplitudes,
\begin{equation} \label{SSolTwin1}
\widehat{b}+  \mu_+ u_+ = \frac{u_+}{\omega\gamma_+},
\end{equation}
\begin{equation} \label{SSolTwin2}
\widehat{b} -  \mu_- u_- =  \frac{u_-}{\omega\gamma_-},
\end{equation}
\begin{equation} \label{SSolME}
\left(k^2 - \omega^2\right) \widehat{b} =  \left(\frac{1}{U_{A0}}\right)^2 \left(u_+ - u_-\right),
\end{equation}
that, for convenience, are displayed in terms of the amplitude of the four velocity $\left(u_\pm=\gamma_{\pm} v_\pm, \gamma_\pm=\sqrt{1+u^2_\pm}\right)$, and $\widehat {b}= b/k$. 

In principle, we can manipulate these equations, treating them as a ``dispersion relation'', i.e. $\omega=\omega(k)$ or $k=k(\omega)$, governing the arbitrary amplitude relativistic CP waves. However, because of the $\gamma$'s, this dispersion relation (``DR'') will be amplitude dependent - a reflection of the highly nonlinear nature of the wave despite its plane wave characteristics. In fact, it is this very amplitude dependence that makes these relativistic waves uniquely interesting; the propagation characteristics are profoundly different from their N.R counterparts.   

Although this system is applicable to any two-component quasi neutral plasma, the ``DR'' becomes especially transparent for a plasma with $\mu_+ = \mu_- = 1/2$ for instance, when we consider an electron-positron plasma with $f_+=f_-=f$; such plasmas are widely studied in both astrophysical and laboratory settings \citep{RVX}. Straightforward manipulations (invoking $\gamma_\pm^2= 1+u_\pm^2$) lead to a quadratic in $\omega^2$, 
\begin{equation}\label{AmpDisp1}
\omega^4 - 4\,\omega^2 \left[\left(\frac{1}{U_{A0}}\right)^2+ \frac{1}{\gamma_+^2 \gamma_-^2} + \frac{k^2}{4}\right] + \frac{4k^2}{\gamma_+^2 \gamma_-^2} = 0,
\end{equation}
which has the exact  solutions
\begin{eqnarray}\label{AmpDisp2}
{\omega_\pm^2} &=& 2\left[\left(\frac{1}{U_{A0}}\right)^2 + \frac{1}{\gamma_+^2 \gamma_-^2} + \frac{k^2}{4}\right] \\
&& \pm 2\left(\left[\left(\frac{1}{U_{A0}}\right)^2 + \frac{1}{\gamma_+^2 \gamma_-^2} + \frac{k^2}{4}\right]^2 - \frac{k^2}{\gamma_+^2 \gamma_-^2}\right)^{1/2}. \nonumber
\end{eqnarray}
In order to explore the essence of the ``DR''  (\ref{AmpDisp1}) and (\ref{AmpDisp2}), let us approximate the higher frequency mode by balancing the first two terms in (\ref{AmpDisp1}),
\begin{equation}\label{AmpDispH}
\omega_{+}^2\approx  4 \left[\left(\frac{1}{U_{A0}}\right)^2+ \frac{1}{\gamma_+^2 \gamma_-^2} + \frac{k^2}{4}\right] ,
\end{equation}
and the lower frequency mode by balancing the second and the third terms 
\begin{equation}\label{AmpDispL}
\omega_{-}^2\approx   \frac{k^2}{\gamma_+^2 \gamma_-^2} \left[{\frac{1}{U_{A0}^2}+ \frac{1}{\gamma_+^2 \gamma_-^2}+\frac{k^2}{4} }\right]^{-1}.
\end{equation}
We can further rewrite these expressions in terms of physical units; (\ref{AmpDispH}) translates to
\begin{equation}\label{AmpDispHPhys}
{\omega_{+}^2}\approx c^2 k^2+ \frac{\omega_{p}^2}{\Gamma_{th}} +\frac{\Omega_c^2}{\Gamma_{th}^2 \gamma_+^2 \gamma_-^2}
\end{equation}
while (\ref{AmpDispL}) becomes 
\begin{equation} \label{AmpDispLPhys}
\omega_-^2 \approx  \frac{k^2 V_{A0}^2}{\Gamma_{th} \gamma_+^2 \gamma_-^2}\left(1 + \Gamma_{th} \delta^2 k^2 + \frac{V_{A0}^2}{c^2} \frac{1}{\Gamma_{th} \gamma_+^2 \gamma_-^2}\right)^{-1},
\end{equation}
where $\omega_p=\sqrt{8\pi n_{0} q^2/m}$ (plasma frequency), $\Omega_c= |q|B_{0}/mc$ (cyclotron frequency), and $V_{A0}= B_0/\sqrt{8\pi n_{0}m}$  (Alfv\'en speed) are defined to be exactly the same as their N.R counterparts for an electron-positron system; note that $\delta= c/\omega_p$ is the electron (or positron) skin depth. The relativistic modifications, contained in the kinematic $\gamma$ and the effective thermal $\Gamma_{th}=f$ factors, have been explicitly displayed. Let us now discuss the new physics that we have uncovered:

1) We begin by recovering the N.R limit $\left(\{f,\gamma_+,\gamma_-\} \rightarrow 1\right)$ in which the higher frequency wave is readily recognized as the upper hybrid-light wave 
\begin{equation}\label{AmpDispHNR}
\omega_{+}^2 = c^2 k^2+ \omega_{p}^2 +\Omega_c^2,  \\\ c^2 k^2= \omega_{+}^2- (\omega_{p}^2 +\Omega_c^2),
\end{equation}
implying that the wave displays real propagation only if the wave frequency $\omega$ exceeds the upper hybrid frequency $\sqrt {\omega_{p}^2 +\Omega_c^2}$. The lower frequency wave is clearly the skin-depth corrected Alfv\'en wave 
\begin{equation} \label{AmpDispLNR}
\omega_-^2 = \frac{k^2 V_{A0}^2}{ \left(1 + \delta^2 k^2 + \frac{V_{A0}^2}{c^2} \right)}.
\end{equation}
2) We shall now demonstrate that the nature of relativistically high amplitude waves in a relativistic plasma is very different from the N.R counterparts. Since we have taken the ambient plasma flow to be zero, relativity modifies the wave ``nature'' through two distinct pathways. By rewriting
(\ref{AmpDispHPhys}) as
\begin{equation}\label{AmpDispHPhysP}
c^2 k^2= {\omega_{+}^2}- \left[\frac{\omega_{p}^2}{\Gamma_{th}} +\frac{\Omega_c^2}{\Gamma_{th}^2 \gamma_+^2 \gamma_-^2}\right] \equiv {\omega_{+}^2}-\omega_\mathrm{cutoff}^2,
\end{equation}
we see that the cut-off frequency (above which the electromagnetic wave can  propagate, i.e. with $k^2>0$), has been drastically reduced:

\noindent a) Relativistically high temperature (if present) decreases $\omega_\mathrm{cutoff}$ by bringing down the contribution of the plasma and cyclotron frequencies by a factor of $\Gamma_{th}^{1/2}$ and $\Gamma_{th}$ respectively. This thermal-induced reduction of the plasma frequency is well-known from previous studies \citep{SB15,MahAs16,HCM17}. To the best of our knowledge, the even stronger thermal reduction of the  cyclotron frequency, although not unexpected (the high temperature electron is naturally heavier), is perhaps being reported for the first time.

\noindent b) An even more drastic reduction of the cutoff frequency is wrought by the relativistic amplitude of the wave that brings it down by a factor $R=\Gamma_{th}\gamma_{+}\gamma_{-}$. This effect turns out to be stronger than expected. We shall discuss this anomalously strong effect further when we impart a relativistic (shear) flow to the plasma.

\noindent c) The relativistically induced effective weakening of the magnetic field is manifested rather eloquently in the Alfv\'enic mode (\ref{AmpDispLPhys}), whose frequency falls well-below - down by the factor $\gamma_{+}\gamma_{-}\sqrt{\Gamma_{th}}$ - the standard N.R Alfven value of $\omega_{N.R}=k V_{A0}= k B_0/\sqrt{8\pi n_{0}m}$ \citep{SK80,AMV09,LAM12}.

3) From a conceptual perspective, the most fascinating property of the amplitude dependent ``DR'', namely (\ref{AmpDisp1}) and subsequent approximations, is that, in the extreme relativistic limit ($\gamma_{+}\gamma_{-}\gg 1$), the amplitude-dependent terms go to zero, and (\ref{AmpDisp1}) becomes a conventional dispersion relation - the propagation of a relativistic amplitude light wave in relativistic plasma embedded in a magnetic field (for $\Gamma_{th}=1$) is determined exactly by the dispersion relation obeyed by a non-relativistic linear wave in an unmagnetized plasma,
\begin{equation}\label{AmpDispHPhysPER}
c^2 k^2= {\omega_{+}^2}-  \frac{\omega_{p}^2}{\Gamma_{th}}. 
\end{equation}
For $\Gamma_{th}\gg1$, the wave increasingly resembles the light wave in vacuum. {\it A magnetized, moderate density, relativistic plasma becomes virtually transparent to high amplitude CP electromagnetic waves; the cutoff frequency falls well below the local plasma or cyclotron frequency.}

4) Although the Alfv\'enic mode could be highly important in studying turbulence in highly relativistic systems, we will not dwell on it here except noting that both of the relativistic effects drive the mode frequency towards zero (i.e. towards much lower frequency).

\section{The inclusion of background flow}
To conform more closely to realistic astrophysical systems (and many laser-created plasmas), and broaden the scope of our enquiry, we repeat the above calculation by including a time independent relativistic shear flow $V_{0}(z)$ along the magnetic field \citep{MMR97,PRM98}; our ansatz has already been introduced in (\ref {AmbV}). The following additional notation will be helpful in appreciating new features introduced by the ambient flow:

1) We will distinguish between two distinct kinematic relativistic factors
\begin{equation}\label{TwoGammas}
\Gamma_{0}^2= \frac{1}{1-V_{0}^2}, \quad  \Gamma_{\pm}^2= \frac{1}{1-V_{0}^2-v_{\pm}^2},
\end{equation}
where $v_{\pm}$, as before, is the velocity amplitude of the wave. The presence of the relativistic flow places stringent upper bound on the velocity wave amplitude since the total speed (${\bf V_{\pm}}\cdot {\bf V_{\pm}}= V_{0}^2+v_{\pm}^2$) is bounded by unity, 
\begin{equation}\label{vLimits}
v^{max}_{\pm}< \sqrt{1-V_{0}^2}= \frac{1}{\Gamma_{0}},
\end{equation}
i.e. the more relativistic the ambient flow, the greater the restrictions on the wave field. 

2) What has to be noted, however, is that even mildly relativistic wave amplitudes can make the ratio
\begin{equation}\label{GamLimits}
G_{\pm}^2= \frac{\Gamma_{\pm}^2}{\Gamma_{0}^2}=\frac{1-V_{0}^2}{1-V_{0}^2-v_{\pm}^2}=\frac{1}{1- \Gamma_{0}^2 v_{\pm}^2} \gg 1 
\end{equation}
For $\Gamma_{0}=100$, a weakly relativistic wave amplitude $v_{\pm}=9.9 \times 10^{-3}$ yields $G_{\pm}^2=100$.

3) The leading order manifestations of the sheared flow are contained in 
\begin{equation}\label{ChangesFlow}
\hat\omega=\omega-kV_{0}, \quad \hat{k}_\pm = k - i \frac{\Gamma_{\pm}^2}{2 L_{sh}},
\end{equation}
where the effective shearing length is defined as $L_{sh}^{-1}= d (\ln V_0)/dz$. For the following treatment to be valid, $kL_{sh}\gg\Gamma_{\pm}$ is required. We will soon demonstrate that the free energy available in the flow shear can amplify the electromagnetic wave as it propagates.

The equivalent amplitude dependent ``DR'' for $V_{0}\ne0$, 
\begin{equation}\label{DRWFlow}
\left(\frac{1}{U_{A0}}\right)^2 \frac{\hat{\omega}^2}{k^2 - \omega^2} = \left(\frac{k}{\hat{k}}\right)^2 \frac{\Gamma_0^2}{\Gamma_+^2 \Gamma_-^2} - \frac{\hat{\omega}^2}{4},
\end{equation}
obtained by essentially replicating the $V_{0}=0$ calculation,
contains both the light wave and the Alfvenic mode, and reduces exactly to (\ref{AmpDisp1}) as $V_0 \rightarrow 0$. Unlike (\ref{AmpDisp1}) however, (\ref{DRWFlow}) is not a quadratic in $\omega^2$ or even $\hat{\omega}^2$. Deferring the discussion of the Alfvenic mode to a forthcoming publication, we write down (in physical units), the approximate "DR"  [$k =k(\omega$)] for the higher frequency light wave 
\begin{eqnarray}\label{lightWaveFlow}
c^2 k^2 &\approx& c^2 k_0^2 - i \frac{\omega^2}{\hat{\omega}^2} \frac{\Omega_{c}^2} {k \widehat{L}_{sh}}; \nonumber \\
c^2 k_0^2 &:=& \omega^2 - \frac{\omega_p^2}{\Gamma_{th}} - \frac{\omega^2}{\hat{\omega}^2} \frac{\Omega_c^2}{\Gamma_{th}^2 }\frac{1}{\Gamma_{0}^2 G_{+}^2 G_{-}^2},
\end{eqnarray}
that, once again, illustrates an enormous reduction in the cutoff frequency. We find that the effective cyclotron frequency is lowered by the factor $R_{Fl}=\Gamma_{th} \Gamma_{0} G_{+} G_{-}$, which, akin to the flow-free reduction factor $R=\Gamma_{th} \gamma_{-}\gamma_{+}$, is anomalously large. The expected values due to the relativistic mass increase (thermal and kinematic) should have been  $\approx \Gamma_{th}\Gamma_{0}$ and $\approx\Gamma_{th} \gamma$ (for $\gamma_{+}\approx\gamma_{-}= \gamma$) respectively in these two cases. The origin of the additional ``enhancement'' factor $G_{+} G_{-}$ ($\gamma$) for $R_{Fl}$ ($R$) will be explored in a more detail in future work; it may occur only in plasmas with equal rest mass components. 

It should be emphasized that, in the flow-free case, the wave amplitudes have to be relativistic ($v_{\pm}$ approaching unity) for $\gamma_{\pm}$, and hence the anomalous reduction factor, to be large. For the plasma with a strongly relativistic background flow ($\Gamma_{0}\gg1$), even very moderate wave amplitudes ($v_{\pm}$ approaching $1/\Gamma_{0}$) are sufficient to make $G_{+} G_{-}$ large, as seen from (\ref{vLimits}) and (\ref{GamLimits}). The latter case, with moderate amplitudes, is more likely to be the pertinent scenario for astrophysical plasmas.

Equation (\ref{lightWaveFlow}) yields an approximate solution for the propagation wave number
\begin{equation}\label{PropNumber}
k \approx k_0 - \frac{i}{2 } \frac{\omega^2}{\hat{\omega}^2} \frac{\Omega_c^2}{k_{0}c^2} \frac{1}{k_{0}\widehat{L}_{sh}}\equiv k_{R}+ik_{I}
\end{equation}
where we have chosen the solution with positive $k_{R}=k_{0}> 0$, i.e, corresponding to an outward propagating wave. $\widehat{L}_{sh}$ can have either sign, since it is set by the sign of $d (\ln V_0)/dz$. Consequently, $k_{I}$ could be either positive or negative, implying that the wave could amplify or grow. Let us substitute (\ref{PropNumber}) into our wave ansatz (\ref{Circv}) and (\ref{Circb}), and note that the phases for the right-handed and left-handed waves (proportional to $\hat{e}_x + i \hat{e}_y$ and $\hat{e}_x - i \hat{e}_y$) are, respectively, 
\begin{eqnarray}\label{PropSFExp}
i(kz-\omega t)=i(k_{0}z-\omega t)- k_{I}z, && \nonumber \\
-i(kz-\omega t)=-i(k_{0}z-\omega t)+ k_{I}z, &&
\end{eqnarray}
thereby demonstrating that one of the two polarizations will always amplify as the wave propagates outwards.

This demonstration of the wave amplification, driven by the free energy in the shear, is extremely important because, without a source of energy, the waves cannot reach significant levels of interest. In future calculations, we intend to present results showing that this class of waves could feed, for instance, on the varying gravitational field in the vicinity of compact objects \citep{MMOC}. It turns out that the difference between the various amplification mechanisms lies only in the specific details. Hence, one could regard the preceding calculation as being generic.

\section{Application of the model}
We end this paper by suggesting how a broad spectrum of electromagnetic waves could escape the magnetosphere of a typical pulsar \citep{KT73,Mich82,ST83}. The relevant electron-positron (e-p) plasma, according to the the standard model \citep{GJ69}, has a representative number density of $10^8$ cm$^{-3}$ (translating to a plasma frequency $\omega_{p}\approx 3\times10^8$ s$^{-1}$), and a streaming $\Gamma_{0}= (1-V_{0}^2)^{-1/2}\approx10^5-10^7$. The energy spectrum has a considerable spread in $\Gamma$ \citep{GJ69}, which, in this simple fluid treatment, will be modeled by an effective temperature measured by $\Gamma_{th}$.
 
An estimate for the magnetic field at the location of the e-p plasma is more challenging. Close to the star's surface, the magnetic field is $1.8\times10^{12} \sqrt {P \dot{P}}$ G where $P$ is the period and $\dot{P}$ has been normalized in units of $10^{-15}$. For the Crab pulsar, we make use of $P=.0332$ s, $\dot{P} =421$, $B_{star}= 6.7\times 10^{12}$ G \citep{BB14}. As the magnetic field falls as $r^3$, the field embedding the e-p plasma - located near the so called light cylinder whose radius $R_{plasma}=c P/2\pi$ \citep{MMOC} - is estimated as $B_0=B_{star} (R_{star}/R_{plasma})^3 \approx 2.5 \times10^6$ G. For a millisecond pulsar with $P=10^{-3}$ s and $\dot{P}=10^3$, the embedding field can be as large as $10^{10}$ G.
 
Let us demonstrate the essence of this work by working with the higher value of $B_0$ for which, in the non-relativistic limit, $\Omega_{c}=2\times 10^{17}$ s$^{-1}$ fully dominates the high cutoff frequency since $\sqrt{\Omega_{c}^2+\omega_{p}^2}\approx \Omega_{c}\approx 2 \times 10^{17}$ s$^{-1}$. The situation changes drastically when we turn on relativistic effects. For a kinematic $\Gamma_{0}=10^6$, a thermal $\Gamma_{0}= 10^3$ (merely $10^{-3}$ of the directed $\Gamma_{0}$), and a $G = 20$ (corresponding to a wave amplitude as small as $.99/\Gamma_{0}$), all relevant frequencies -- the effective cyclotron, plasma, and consequently the cutoff frequencies -- approach $10^7$ MHz, i.e. a remarkable fall by $10$ orders of magnitude. The pulsar magnetosphere of a typical millisecond pulsar is rendered transparent to moderate amplitude electromagnetic waves with frequency greater than $\sim 10$ MHz. If there existed a suitable energy source (eventually gravity) to amplify waves to such levels (amplification ends at the plasma-vacuum boundary), they could propagate outwards and be observed by an Earth-based detector.

We have neglected the role of nonlinear quantum electrodynamics (QED) in our analysis; although these effects are potentially important in pulsar magnetospheres \citep{BGI93,CB17}, most studies do not take them into consideration. For instance, it is possible that nonlinear QED contributions can counteract the onset of relativistic transparency for ultra-relativistic plasmas \citep{ZRP15}.

\section{Conclusion}
The potentially new results derived in this paper -  that the large amplitude CP waves propagating in highly relativistic plasmas embedded in a magnetic field behave like linear waves in a non-relativistic system - should find applications in a variety of astrophysical and high energy laser-plasma settings. The first part of this paper (without an explicit source of free energy) made a powerful conceptual statement that such  waves, with well-defined characteristics, were sustainable in a relativistic two-fluid plasma. But, it was the second part (with the inclusion of a finite shear flow) that delineated a concrete pathway for these waves to attain large amplitudes.

\acknowledgments
The work of S.M. was partly supported by U.S. DOE Contract No. DE-FG. 03-96ER-54366.


%

\end{document}